\documentclass[prl,twocolumn,aps,superscriptaddress,nofootinbib]{revtex4-2}
\usepackage{amsfonts}
\usepackage{mathtools} % mathtools is an upgrade of amsmath
\usepackage[svgnames,table]{xcolor}  % xcolor > color
\usepackage[T1]{fontenc}
\usepackage{txfonts}

\usepackage{graphicx}

\usepackage{hyperref}
\hypersetup{
        colorlinks=true,
        citecolor=SteelBlue,
        filecolor=LimeGreen,
        linkcolor=SlateBlue,
        urlcolor=MediumPurple
}

\newcommand{\yanbei}[1]{\textcolor{RoyalBlue}{#1}}
\newcommand{\Xiang}[1]{\textcolor{Green}{#1}}

\newcommand{\be}{\begin{equation}}
\newcommand{\ee}{\end{equation}}

\graphicspath{{Figures/}}

\begin{document}
\title{Broadband sensitivity improvement via coherent quantum feedback with PT symmetry}

\author{Xiang Li}
\affiliation{Burke Institute for Theoretical Physics, California Institute of Technology, Pasadena, California}

%\author{Yifan Chen}
%\affiliation{Institute of Theoretical Physics, Chinese Academy of Sciences, Beijing, China}
\author{Maxim Goryachev}
\affiliation{ARC Centre of Excellence for Engineered Quantum Systems and ARC Centre of Excellence for Dark Matter Particle Physics, Department of Physics, University of Western Australia, Crawley,WA, Australia.}
\author{Yiqiu Ma}
\affiliation{Burke Institute for Theoretical Physics, California Institute of Technology, Pasadena, California}
%\affiliation{Department of physics, Huazhong University of Science and Technology, Wuhan, China}
%\author{Jing Shu}
%\affiliation{Institute of Theoretical Physics, Chinese Academy of Sciences, Beijing, China}
\author{Michael E. Tobar}
\affiliation{ARC Centre of Excellence for Engineered Quantum Systems, University of Western Australia, Crawley, WA, Australia}
%\author{Robert Ward}
%\affiliation{Australian National University, Canberra, ACT, Australia}
\author{Chunnong Zhao}
\affiliation{OzGrav, University of Western Australia, Crawley, WA, Australia}
\author{Rana X Adhikari}
\affiliation{Bridge Laboratory of Physics, California Institute of Technology, Pasadena, California}
%\author{Vaishali Adya}
%\affiliation{Australian National University, Canberra, ACT, Australia}
\author{Yanbei Chen}
\affiliation{Burke Institute for Theoretical Physics, California Institute of Technology, Pasadena, California}

\begin{abstract}
  A conventional resonant detector is often subject to a trade-off between bandwidth and peak sensitivity that can be traced back to quantum Cramer-Rao Bound. 
  %This limitation can be traced back to the quantum Cramer-Rao Bound, and the fact that the quantum state of the resonator is often in a coherent state.
  %The trade-off can be avoided by injecting non-classical quantum states (e.g., squeezed vacuum), but such quantum-noise suppression strategies often have stringent requirements on optical losses.
  Anomalous dispersion has been shown to %be able to  %a detector's bandwidth-sensitivity 
  improve it by signal amplification %[Phys. Rev. D 99, 102001 (2019)] 
  and is thus more robust against decoherence, while it leads to instabilities. %In standard linear systems, however, the addition of anomalous dispersion can lead to instabilities.
  We propose a stable quantum amplifier applicable to linear systems operating at the fundamental detection limits, %spectrum expander
  enabled by two-mode non-degenerate parametric amplification. At threshold, one mode of the amplifier forms a $\mathcal{PT}$-symmetric system of original detector mode. %, while the other mode of the amplifier collects signal and gets extracted by a readout.
  %Viewed more broadly in the context of coherent quantum control theory: we are attaching a  {\it controller} that consists of a time reversal of the {\it plant}, which, by canceling the inertia of the plant, helps drive up the {\it closed-loop} signal gain.
  %This technique is generally applicable to linear systems operating at the fundamental limits of detection. 
  Sensitivity improvements are shown for laser-interferometric gravitational-wave detectors and microwave cavity axion detectors.
\end{abstract}

\maketitle

\noindent {\it Introduction.---}
Oscillators are often used to measure weak classical forces. In the early days of gravitational-wave (GW) detection,  excitations of mechanical oscillators (resonant bars) %large bars
%near their eigenfrequencies
were read out by inductive, capacitive\,\cite{solomonson1994construction} or parametric\,\cite{Blair1995High} transducers  %\Xiang{[not sure about the historical discussion trend, according to Ref.\,\cite{solomonson1994construction}, it seems that both capacitive and inductive transducers were proposed around 1976, shall we mention both of them here? [MET] Parametric (or optomechanical) transducers were used as well \cite{Blair1995High}: DG Blair, EN Ivanov, ME Tobar, PJ Turner F Van Kann and IS Heng, "High sensitivity gravitational wave antenna with parametric transducer readout," Phys. Rev. Lett., vol. 74, no. 11, pp. 1908-1911, 1995 ]} % sensors/ transducer takes measurement in one form converts it to another
and via a superconducting quantum
interference device (SQUID) amplifer\,\cite{solomonson1994construction,Blair1995High,Mauceli1996Allegro}.  %later, the optical modes of km-scale Fabry-Perot cavities are excited and
A more sensitive technique is to read out the phase fluctuations in a light beam (Michelson and Fabry-Perot type interferometers)~\cite{abbott2004detector,adhikari2014gravitational,Abbott2016Observation} using photodetection.
Signal recycling~\cite{meers1988recycling} and Resonant Sideband Extraction (RSE)~\cite{mizuno1993resonant} techniques were designed optimize GW sensitivity by tailoring the optical frequency response. In this process, Mizuno noticed a trade off between bandwidth and peak sensitivity~\cite{mizuno1995comparison} -- analogous to the gain-bandwidth product in electronic amplifiers~\cite{Faulkner1970Active}.
%\Xiang{(the earliest literature I can find)}. Or we can cite some textbook
Braginsky {\it et al.}~\cite{braginsky1995quantum,braginsky2000energetic} showed\,\footnote{More specifically, they assumed a single cavity optimally aligned with an incoming plane GW with wavelength much larger than the cavity length.}, using the energy-phase uncertainty relation, that the power spectral density of equivalent spacetime strain noise is
%\begin{equation}
$S_h (\Omega)   \ge 4\hbar^2/S_\mathcal{E}(\Omega) $
%\end{equation}
where $S_{\mathcal{E}}$ is the spectral density of energy in the cavity, and
\begin{equation}
\label{eq:GBWproduct}
\int_{0}^{+\infty} d\Omega/(2\pi) S_h^{-1}(\Omega) \le  \Delta\mathcal{E}^2/(4\hbar^2)\,.
\end{equation}
This integral embodies a {\it bandwidth-sensitivity trade-off}. This was also obtained by Tsang, et al.\ using Quantum Fisher Information~\cite{tsang2011fundamental}, and further elaborated in Refs.~\cite{miao2017towards,pang2018quantum,pang2019fundamental,miao2019quantum}.
%For a classical signal, $x$, coupled to the device operator $\hat F$,
%\begin{equation}
%\label{eqQCRB}
  %  \hat V = - x(t) \hat F(t)\; \Rightarrow\; S_{xx} \ge S_{xx}^{\rm QCRB} = \hbar^2/S_{FF}\,,
%\end{equation}
%where QCRB stands for the Quantum Cramer-Rao Bound.  
For coherent states, $\Delta \mathcal{E}^2=\hbar \omega_0 \mathcal{E}$, and in this case Eq.~\eqref{eq:GBWproduct} is also referred to as the  {\it Energetic Quantum Limit} (EQL) for GW
detection~\footnote{ Refs.~\cite{braginsky1995quantum,braginsky2000energetic,tsang2011fundamental,miao2017towards,pang2018quantum,pang2019fundamental} all used the double-sided definition for spectral density, while in this paper, in Eq.~\eqref{eq:GBWproduct},  we have used the single-sided definition in order to be consistent with most GW literature~\cite{buonanno2001quantum,buonanno2002signal}. }.
The EQL trade-off applies to all quantum metrology experiments that use oscillators at coherent states.

%\yanbei{Mention Bode-Fano?}\Rana{more deets plz} \yanbei{Bode-Fano bound was cited by the paper by Chaudhuri, and this is probably the same as the EQL. Perhaps Xiang can read the paper by Chaudhuri to find out how they had used the B-F bound. }

\noindent{\it Detectors.---}  In this paper, we consider two types of detectors. 
%\yanbei{Here we insert brief physics discussions on how axion detectors and GW detectors work.}
GW detectors (of the laser interferometer type~\cite{Abbott2016Observation}) use optical resonators to increase the
interaction between the spacetime strain and the laser light field. Axion detectors~\cite{ADMX:2020} (of the Sikivie Haloscope~\cite{Sikivie:2013} type) use microwave resonators and a powerful permanent magnet to increase the interaction with the axion field and the microwave field. In both cases, the emphasis is on increasing the quality factor~\cite{Stern:2015eu}, $Q$, of the resonance so as to maximize the transduction coefficient between the physical signal and an electrical readout variable. However, the EQL limits the useful bandwidth of the detector according to the gain-bandwidth tradeoff in Eq.~\eqref{eq:GBWproduct}.

In GW detection, the EQL can be surpassed when non-classical states of light are created for the arm cavity, with $\Delta \mathcal{E}^2 > \mathcal{E}^2/\bar N$~\cite{zhou2015quantum,miao2015enhancing,korobko2017beating,korobko2019quantum}. This can be implemented via squeezing injection~\cite{kimble2001conversion}, or internal ponderomotive squeezing achieved by optical springs~\cite{buonanno2001quantum,buonanno2002signal,miao2017towards}. To directly address the EQL, another concept of White-Light Cavity (WLC), which resonates with a broader spectrum of frequencies without sacrificing sensitivity, has been proposed~\cite{wicht1997white,wise2004linewidth,wise2005phase,pati2007demonstration,ma2015quantum,zhou2015quantum,miao2015enhancing,korobko2017beating,korobko2019quantum}.
%This is only possible when non-classical states of light are created for the arm cavity, with $\Delta \mathcal{E}^2 > \mathcal{E}^2/\bar N$~\cite{zhou2015quantum,miao2015enhancing,korobko2017beating,korobko2019quantum}.
Despite beating the EQL in theory, in some cases~\cite{korobko2017beating,korobko2019quantum}, improving sensitivity by suppressing noise, while keeping signal at the same level, makes the system more susceptible to optical losses --- similar to the situation when injecting squeezed vacuum~\cite{caves1981quantum,kimble2001conversion}.

\begin{figure}
\includegraphics[width=\columnwidth]{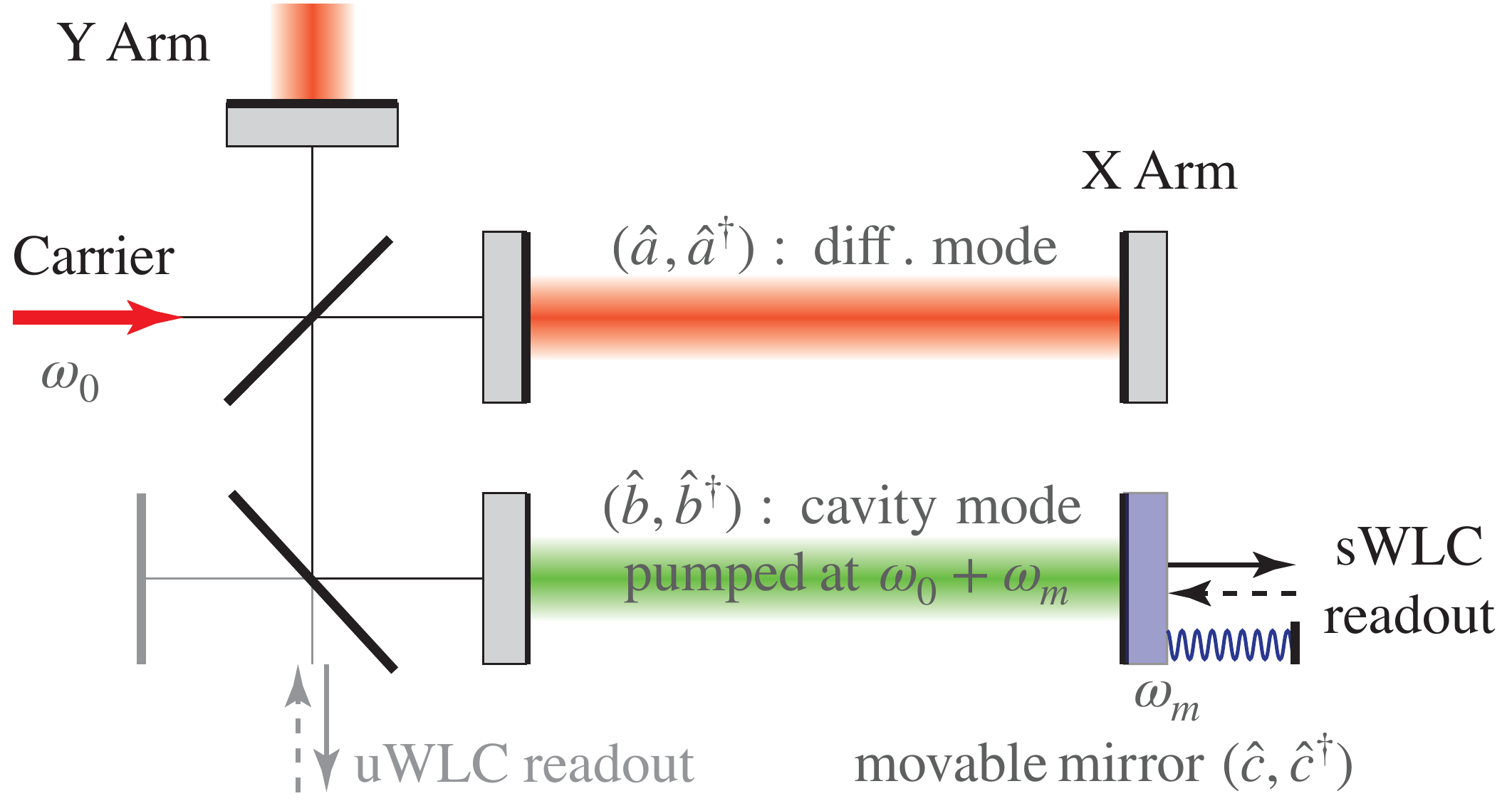}
\caption{Coherent quantum feedback for laser interferometer GW detectors.   In the uWLC scheme (signal readout shown in gray), positive storage time in the arms is canceled by anomalous dispersion of the filter cavity, which has a movable mirror with mechanical oscillation frequency $\omega_m$ and is pumped at $\omega_0+\omega_m$.   The {\it controller}  is the time reversal of
  the {\it plant}.  The sWLC scheme extracts signal at the end of the filter cavity (shown in black).  Both schemes are also illustrated in Fig.~\ref{fig:hamiltonian}.
\label{fig:uWLC}}
\end{figure}

\begin{figure}
\includegraphics[width=\columnwidth]{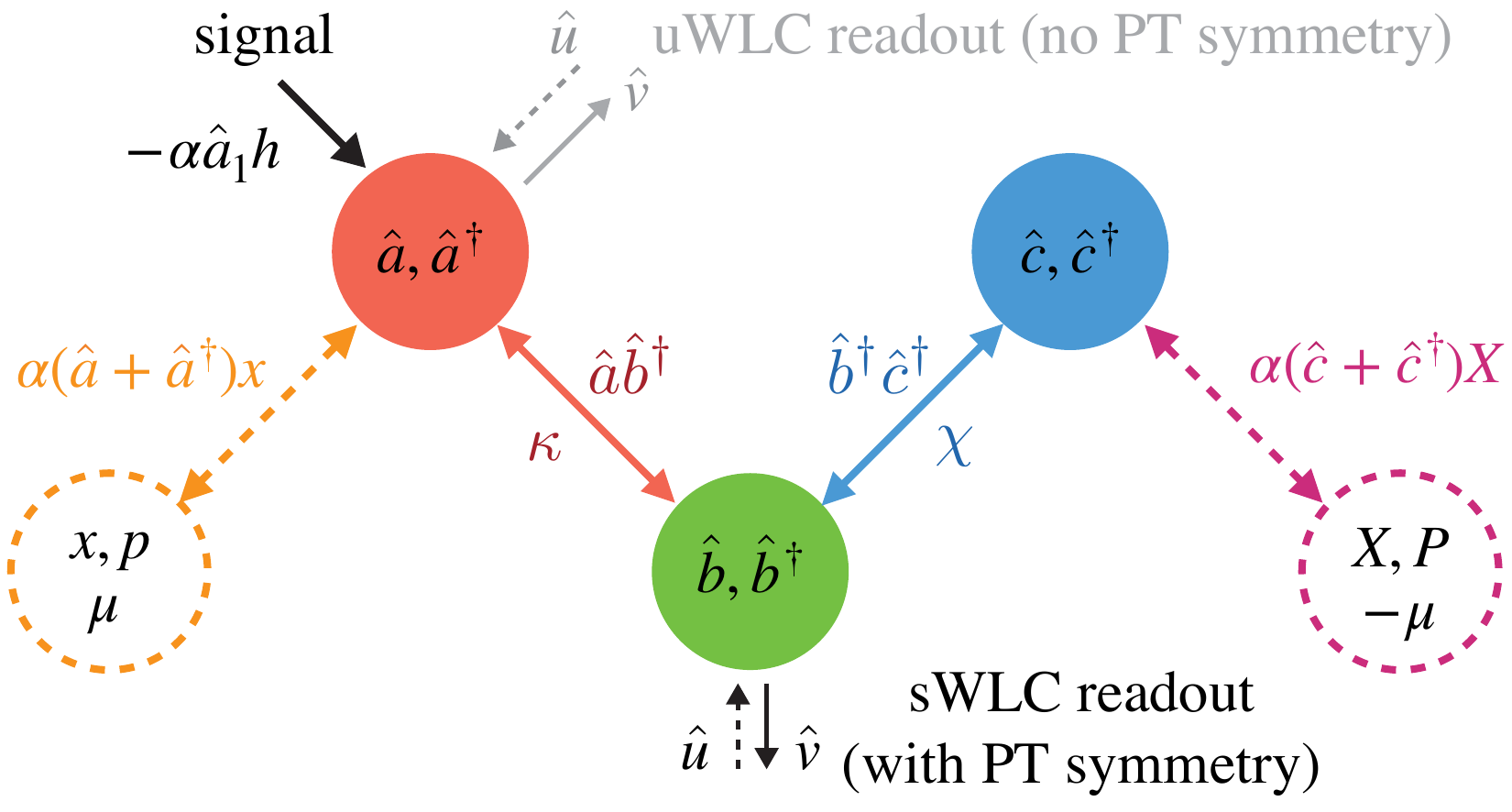}
\caption{A force sensor [mode $(\hat a,\hat a^\dagger)$, plus $(x,p)$ for the test mass in case of GW detectors] plus coherent quantum feedback with modes $(\hat b,\hat b^\dagger)$, $(\hat c,\hat c^\dagger)$ and $(X,P$). Instead of directly coupling to readout, the mode $\hat a$ first couples to a mode $\hat b$ (via $\hat a^\dagger \hat b$), which in turn couples to $\hat c$ (via $\hat b^\dagger \hat c^\dagger$).  Mode $\hat c$ may also couple to $(X,P)$ with negative mass {to cancel the back-action noise caused by $\hat a$'s coupling to $(x,p)$ (back-action evasion, see Fig.~\ref{fig:GWspectra}).} %\Xiang{, i.e. canceling the measurement induced back-action} %canceling the radiation pressure noise exerted on the test mass.} %The $\hat a$-$\hat b$ coupling is of the beam-splitter type, while the $\hat b$-$\hat c$ coupling is of the parametric amplifier type. 
The mode $\hat b$ is coupled to an external continuum (``sWLC readout'' in black). System is $\mathcal{PT}$ symmetric when $\kappa=\chi$. The uWLC scheme is recovered when $\hat a$ couples to continuum instead (shown in gray); further setting $\kappa\rightarrow 0$ reduces to a conventional detector.
%For general multi-mode sensors with $\{\hat a_j\}$, we need an auxiliary multi-mode system with $\{\hat c_j^\dagger\}$ that satisfy the same equations of motion.
% Insert: mode interaction structure for arm cavity readout scheme in Ref\,\cite{miao2015enhancing} (Fig.\,\ref{fig:uWLC}), instead of the filter cavity mode $\hat b$, the original cavity mode $\hat a$ directly couples to the continuum. That configuration breaks the $\mathcal{P}$ symmetry and thus doesn't have the fancy features discussed in this paper. 
%\Rana{[replace with a cool 3D image]}
%\yanbei{New conventions: $g\rightarrow \kappa$, $G\rightarrow \chi$.\, $\gamma$ to $\gamma_R$.} \Xiang{add $\hat{}$ to the mode operators in the FIG}
}
\label{fig:hamiltonian}
\end{figure}

In the unstable WLC (uWLC) design of Miao {\it et al.}~\cite{miao2015enhancing} (Fig.~\ref{fig:uWLC}), an  additional {\it coherent quantum feedback controller}~\cite{mabuchi2005principles,nurdin2009coherent} (enabled by an {\it unstable} optomechanical oscillator) is attached to a laser interfereomter to provide an ``anomalous dispersion'' whose {\it negative group time delay} cancels the positive group time delay in the interferometer, achieving broadband signal amplification without increasing noise. The instability of this quantum system can, in principle, be stabilized by a classical controller without adding noise~\cite{buonanno2001quantum,buonanno2002signal} %\Rana{cite optical spring experiments that proved this:40m, Corbitt}.
%Since the anomalous dispersion is a {\it time reversal} of the ordinary dispersion, the uWLC is $\mathcal{T}$-symmetric.
%\Xiang{However, this anomalous dispersion causes a dynamical instability that needs to be stabilized by a classical feedback loop.}
%, which increases technical difficulty and conceptual complexity.}

In this paper we introduce a parity-time ($\mathcal{PT}$)-symmetric coherent quantum strategy~\cite{liu2016metrology,Bentley2019Converting} that not only leads to a stable WLC (sWLC), but also applies to a wide range of quantum systems.
%
%a stable signal amplification strategy based on $\mathcal{PT}$-symmetric coherent quantum control. We will introduce {\it parity reversal} symmetry in addition to the {\it time reversal} feature of anomalous dispersion, by modifying how the modes interact. 
%
%
%A similar amplification strategy in the classical domain has been proposed for axion detection~\cite{goryachev2019axion,goryachev2019probing}. 
We first consider the simplest sWLC, showing that it approaches an {\it Exceptional Point} (EP)\,\cite{Wiersig2016Sensors,Zhang2019Quantum} as the feedback gain reaches the threshold of stability, at which point the theoretical gain in sensitivity is infinite.  We then show that the design strategy is very general, since the controller's Hamiltonian always corresponds to the time reversal of that of the plant.  We finally discuss applications to GW and axion detection.

\noindent {\it Stable Optomechanical Amplifer.---}
Suppose the mode of a weak-force sensor has annihilation/creation operators $(\hat a,\hat a^\dagger)$,
which can be reorganized into quadrature operators $\hat a_1 =(\hat a+\hat a^\dagger)/\sqrt{2}$, $\hat a_2 =(\hat a-\hat a^\dagger)/(\sqrt{2} i)$,
and the detector is  coupled to a signal $h$ via $\hat V_s = -\alpha \hat  a_1 h$.
For a GW detector, $\hat a$ is the cavity mode, $h$ is GW strain,
while $\alpha$ is achieved by strong carrier field in the cavity (we first ignore
radiation-pressure effects~\cite{kimble2001conversion}; $h$ will be replaced by the axion field $\Psi$ later in the paper). 

\begin{figure}
  \includegraphics[width=\columnwidth]{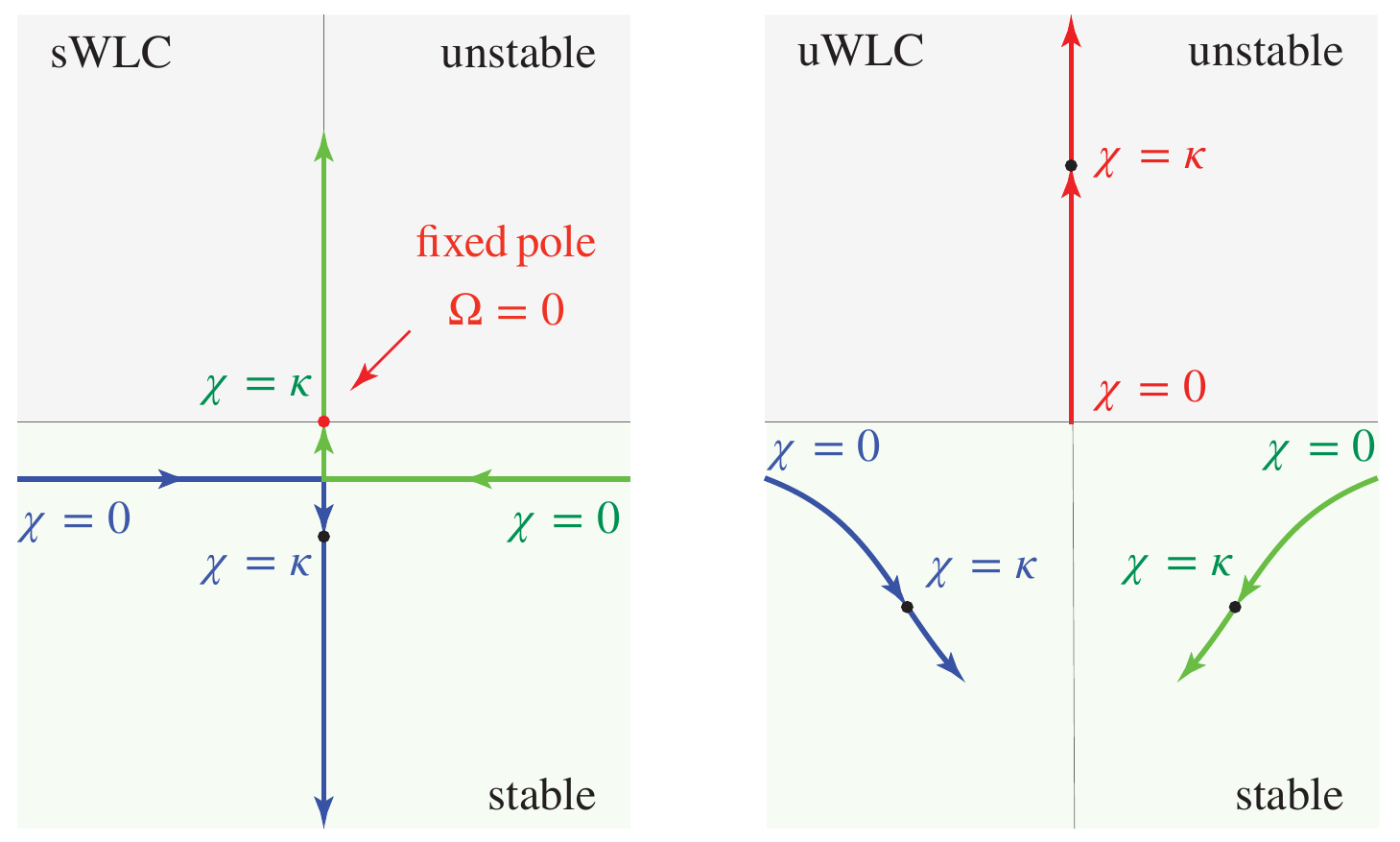}
  \caption{Trajectory of poles of WLC dynamics in the complex plane (dashed curves), with direction of increasing $\chi$ (with $\kappa$ fixed) indicated by arrows. Left: the sWLC has a fixed pole at $\Omega=0$ (independent from $\chi$), which is invisible to the input-output relation~\eqref{eq:v2amp}; the other two poles first move toward the imaginary axis and meet each other, then move up and down separately, until one of them reaches $0$ when $\chi=\kappa$. For $\chi>\kappa$, the system becomes unstable. Right: on contrary, the uWLC is unstable as soon as $\chi>0$. 
}
\label{fig:poles}
\end{figure}

As in Fig.~\ref{fig:hamiltonian}, let us couple $\hat a$ to $\hat b$ with rate $\kappa$, while $\hat b$ and $\hat c$ are together amplified by a non-degenerate amplifier with rate $\chi$,
the Hamiltonian
reads (Sec.~2 of Ref.~\cite{chen2013macroscopic}):
\begin{equation}
\label{eq:vamp}
\hat V =  i\hbar\kappa (\hat a \hat b^\dagger -\hat a^\dagger \hat b) + i \hbar\chi (\hat b^\dagger \hat c^\dagger -\hat b\hat c).
\end{equation}
The mode $\hat b$ is read out by being coupled to a continuum $\hat w_\xi$ of
incoming $(\xi<0)$ and out-going $(\xi>0)$ fields at $\xi=0$, with
$\hat V_b =  i \sqrt{2\gamma_R} \hat  b \hat w^\dagger_{\xi=0} +\int_{-\infty}^{+\infty} d\xi\left[ i\hat  w^\dagger _\xi \partial_\xi \hat w_\xi  +\mathrm{h.c.}\right]$,
and we shall use $\hat u$ and $\hat v$ to denote $\hat w_{0-}$ and $\hat w_{0+}$,
the field right before entering and right after leaving.
(In uWLC the operator $\hat a$, instead of $\hat b$, is coupled to the continuum $\hat w_\xi$.)
The resulting Heisenberg equations are given by
\begin{align}
\label{eq:Heqqamp}
\dot {\hat  a}   =   i \alpha h-\kappa\hat b ,\; \dot {\hat  c}^\dagger  = \chi \hat b,\;
\dot {\hat b}  = -\gamma_R \hat b + \kappa\hat  a + \chi\hat  c^\dagger +\sqrt{2\gamma_R}\hat  u, 
\end{align}
and $\hat v  =\hat  u -\sqrt{2\gamma_R}\hat  b$. The phase quadrature of the output field reads:
\begin{equation}
\label{eq:v2amp}
\hat v_2 =\frac{\Omega^2-i\gamma_R\Omega+\chi^2-\kappa^2}{\Omega^2+i\gamma_R\Omega+\chi^2-\kappa^2}\hat u_2 + \frac{2\sqrt{\gamma_R}\kappa\alpha  }{\Omega^2+i\gamma_R\Omega+\chi^2-\kappa^2}h\,.
\end{equation}
From poles of $\hat v_2$, we see that increasing the amplification rate $\chi$ tends to destabilize the system, while increasing $\kappa$ tends to stabilize it; the entire system is stable when $\chi\leq\kappa$. The system also has a fixed pole at $\Omega=0$ not visible in $\hat v_2$.  We plot trajectories of poles for sWLC and uWLC dynamics in Fig.~\ref{fig:poles}. 

%The output fields have two poles, which are {\it stable} as long as  $G\le g$, while the whole system has another fixed pole at $\Omega =0$. In Fig.~, we plot trajectories of these poles in the complex plane, and compare with the uWLC case.  

\begin{figure}
\includegraphics[width=\columnwidth]{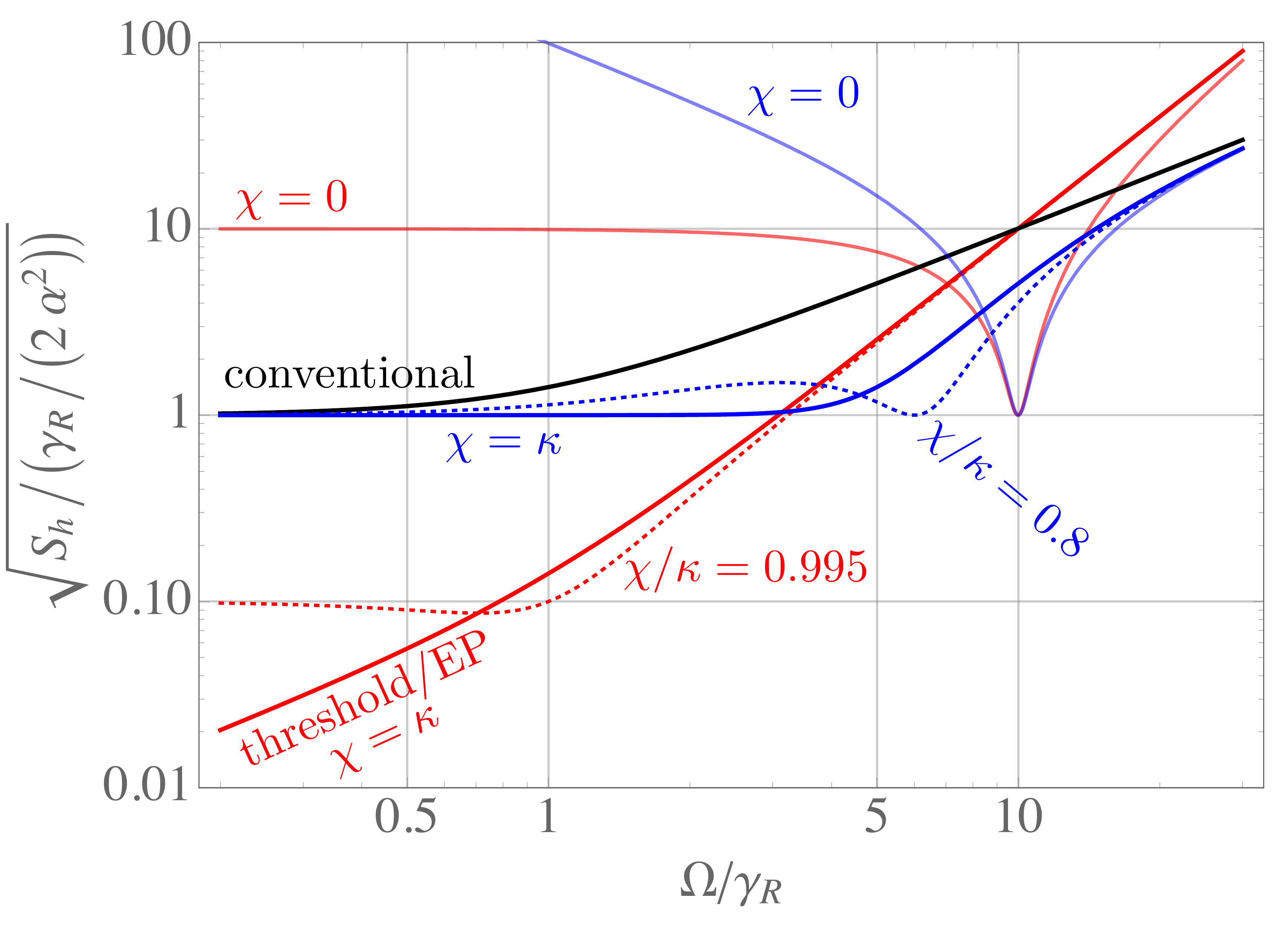}
\caption{\label{eq:idealWLCs}Coherent amplification of sensitivity by sWLC (red) and uWLC (blue) when $\kappa=10\gamma_R$ for a lossless single-mode detector; Connection between threshold and EP can be found in the main text. For the same choice of $\gamma_R$ and $\chi$, uWLC tends to improve higher frequency sensitivity, while sWLC tends to improve lower frequency sensitivity. %\Xiang{Although sWLC is more vulnerable to optical and mechanical mode losses, it is still more promising when BAE technique is applied, see FIG.\,\ref{fig:NoiseBudget} and the main text for more details.} %its optimized integrated sensitivity is still better than the optimized case of sWLC, see Appendix for more details.} %\yanbei{[Erase some traces. 
%\Xiang{[will the baseline case mentioned in appendix still be useful?]}
}
\label{fig:curves}
\end{figure}

In the single-cavity (or conventional) detector 
(i.e., $\hat a$ is read out directly, without adding $\hat b$ and $\hat c$) has:
%\begin{equation}
%\label{v2conv}
$\hat v_2 =({\Omega-i\gamma_R})/({\Omega+i\gamma_R})\hat u_2 - 2i \sqrt{\gamma_R}\alpha h  /({\Omega+i\gamma_R})$.
%\end{equation}
For vacuum fluctuations, we have the spectrum $S_{\hat u_i \hat u_j}=\delta_{ij}$, and thus
\begin{equation}
\label{eq:spectrumwlc}
S_h^{\rm con} = \frac{\Omega^2+\gamma_R^2}{2\gamma_R\alpha^2},\;
S_h^{\rm amp} = \frac{(\Omega^2-\kappa^2+\chi^2)^2 +\gamma_R^2\Omega^2 }{2\gamma_R \kappa^2\alpha^2}.
\end{equation}
A conventional detector has
$\int_0^{+\infty}  d\Omega/(2\pi) [S_h^{\rm con}]^{-1}  =  \alpha^2/2
$
and the gain in integrated sensitivity by amplification is
\begin{equation}
\label{gaindef}
\Lambda \equiv \frac{\int_0^{+\infty}  d\Omega/(2\pi) [S_h^{\rm amp}]^{-1} }{\int_0^{+\infty}  d\Omega/(2\pi) [S_h^{\rm conv}]^{-1} } = (1-\chi^2/\kappa^2)^{-1}
\end{equation}
with $\Lambda\rightarrow +\infty$ as $\chi\rightarrow \kappa\Xiang{-}$. %\Xiang{(Xiang: should we also compare the integral with uWLC here?)} 
Both configurations have the same noise level in $\hat v_2$, while the amplifier has higher signal amplitude, especially as $\chi \rightarrow \kappa\Xiang{-}$.
 The broadband improvement for the case $g=10\gamma$ is shown in Fig.~\ref{fig:curves}.
 
 %In the concrete case of $G^2=g^2-\gamma^2$, the detection bandwidth will be $\gamma$, and DC noise spectrum of $S_h(0) = \gamma^3/2(g^2\alpha^2)$, with an integrated-sensitivity gain of $g^2/\gamma^2$ from conventional detector.

%Old caption below
 % (a) The two \emph{external} poles of sWLC in Eq.\,\eqref{}: one is represented by the green line and is always inside the stable regime; the other is represented by the blue line in the stable regime when $G<g$ but is continued by the red line in the unstable regime when $G>g$. Additionally, there is another pole inside the internal dynamics at $(0,0)$ that is invisible to the input-output relation. Thus, the overall system meets an exceptional point when $G=g$. Note that the two external poles meet at another exceptional point, pictured by the green dot, when $G=\sqrt{g^{2}-\gamma^{2} / 4}$, but it contributes nothing special in the spectrum. 
  %
%  (b) The three poles of uWLC in Eq.\,\eqref{uWLCsol} (Ref.~\cite{miao2015enhancing}): two poles inside the stable regime are represented by the blue and blue lines separately; however, the other pole represented in red is always inside the unstable regime which contributes to the fact that uWLC can never be inherently stable.

\noindent {\it Effects of Decoherence.---}  Suppose the ($\hat a$, $\hat b$, $\hat c$) modes have dissipation rates $\gamma_a$, $\gamma_b$ and $\gamma_c$, respectively, to their baths, and have noises $\hat n_a^L$, $\hat n_b^L$, $\hat n_c^L$ entering, we need to apply
%\Xiang{(Xiang: avoid double usage of $\hat X$ operator - and please check overleaf logs for red/orange errors)}
\begin{equation}
\label{eq:loss}
\dot {\hat Y}=... \rightarrow \dot {\hat Y}=... -\gamma_Y {\hat Y}+\sqrt{2\gamma_Y} \hat n_Y^L,
\end{equation}
with $Y=a,b,c$ to Eq.~\eqref{eq:Heqqamp}. Depending on the different nature of noise sources, the decay rate $\gamma_Y$ and noise spectrum $S_Y^L$ vary.  %Depending on the different noise sources of a specific kind of detectors, 
%The decay rate $\gamma_Y$ and noise spectrum $S_Y^L$ depend on the nature of baths for a specific kind of detectors. 
\footnote{See Supplemental Material\,\cite{} for more details on decoherence.}

%[mention later on detector sections] \Xiang{For the laser interferometer type GW detectors, $\hat a,\,\hat b$ modes suffer from optical loss with vacuum bath spectrum $S_{a,b}^L=1$, while the $\hat c$ mode decays to a thermal bath with spectrum depending on environmental temperature\,\cite{ma2014narrowing,miao2015enhancing}. The details can be found in the Supplemental Material\,\cite{}. For Sikivie Haloscope type axion detectors, thermal, noise temp of current [????what is its noise source????]}

%We obtain additional noise contributions
%\begin{equation}
%v_2^L =\frac{2\sqrt{\gamma} [g\sqrt{\gamma_a} n_2^a -i\sqrt{\gamma_b}\Omega %n_2^b-2G\sqrt{\gamma_c} n_2^c]}{\Omega^2+i\gamma\Omega+G^2-g^2}
%\end{equation}

\noindent {\it Connection to $\mathcal{PT}$ symmetry and generalization to multi-mode systems.}---  On
%The dynamical system described by Eq.~\eqref{eq:Heqqamp} has 3 poles, the input-output relation in Eq.~\eqref{eq:v2amp} only contains 2; the pole at $\Omega=0$ does not appear. \Xiang{The locations of poles in the input-output relation and their changings are plotted in Fig.\,\ref{fig:poles}.}
threshold at $\chi=\kappa$, poles of the output field $\hat v_2$ are at $\Omega=0,-i\gamma_R$, making $\Omega=0$ a double pole (see Fig.~\ref{fig:poles}). At this point, the system only has two independent eigenmodes, making it an Exceptional Point (EP)\,\cite{Wiersig2016Sensors,Zhang2019Quantum}. 
The system is also $\mathcal{PT}$ symmetric: the Hamiltonian is invariant if exchanging $\hat a\rightarrow \hat c^\dagger$ and $\hat c\rightarrow \hat a^\dagger$.  Specifically, the $\mathcal{P}$ (parity) transformation swaps the $(\hat a,\hat a^\dagger)$-mode and the $(\hat c,\hat c^\dagger)$-mode,  the $\mathcal{T}$ (time-reversal) transformation swaps  creation and annihilation operators; combination of $\mathcal{PT}$ transformation leaves $\hat b$ invariant.
%\yanbei{\sl Yanbei: we need to add root locus; perhaps for both uWLC and sWLC.}

\begin{figure}
\includegraphics[width=\columnwidth]{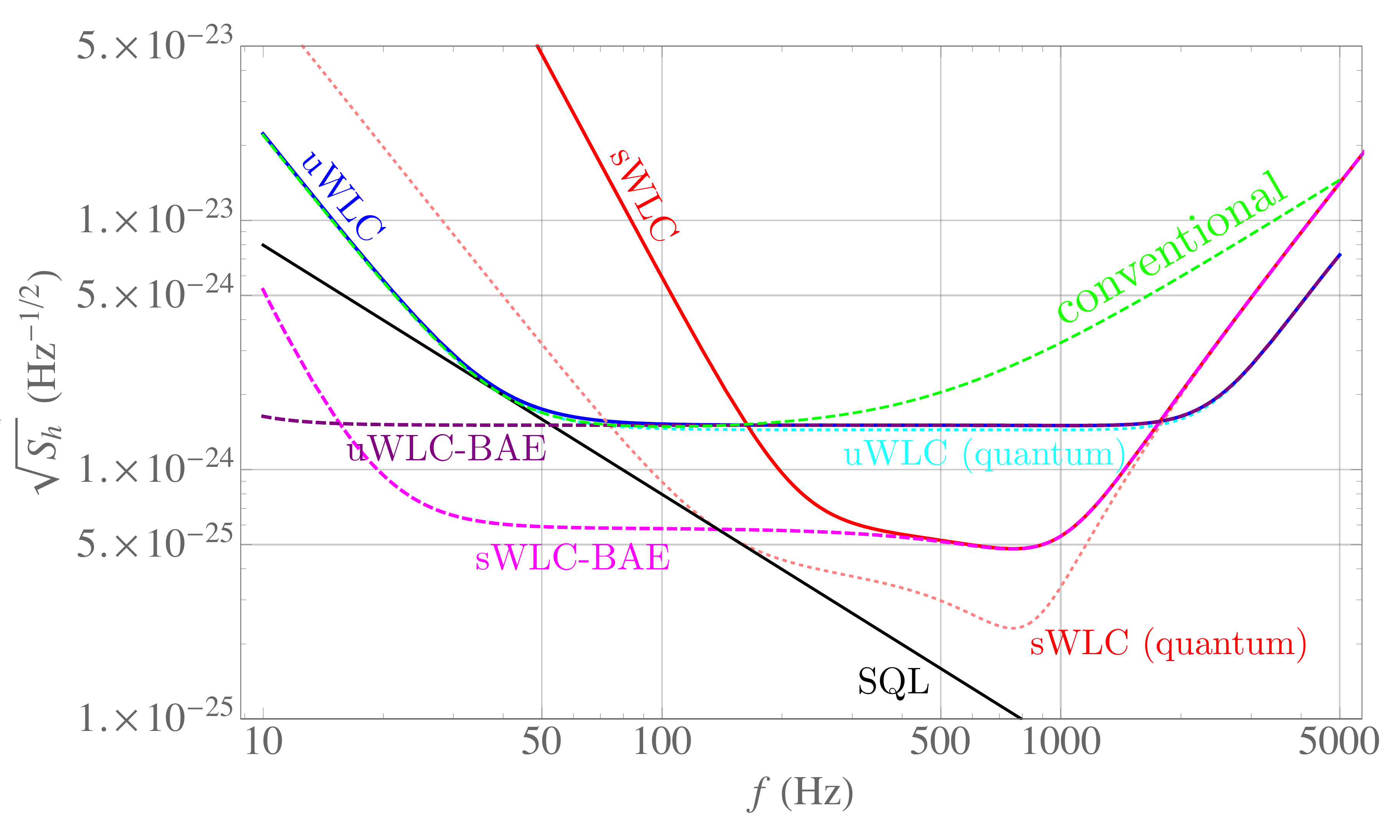}
\caption{\label{fig:GWspectra}
Example of GW noise spectra for sWLC and uWLC (both with $\kappa/(2\pi)=5\,\rm{kHz},\ \chi/(2\pi)=4.93\,\rm{kHz}$) configurations, {\it only} accounting for quantum noise and thermal noise from the mechanical oscillator $\hat c$. Solid red and blue curves are sWLC and uWLC (with dotted pink and cyan curves showing the quantum noise contribution), while dashed magenta and purple curves are BAE configurations that require additional nonlinear elements. Shown for comparison is the quantum noise spectrum for the conventional single cavity (using tuned RSE configuration) without optical loss. Note that in this figure, sWLC, uWLC and conventional are plotted with the same carrier wavelength, arm cavity power, $\hat c$-mode mechanical loss (only for WLC), thermal bath condition (environment temperature), and the readout rate $\gamma_R/(2\pi)=500\,\rm{Hz}$. %[different $\chi,\kappa$?? - use the same $\kappa$]}
%[old caption] GW noise spectra achievable by sWLC and uWLC configurations, {\it only} account for quantum noise and thermal noise from the mechanical oscillator $\hat c$. Solid red and blue curves are sWLC and uWLC (dotted pink and cyan curves show quantum noise contribution), while dashed magenta and purple curves are BAE configurations that require additional nonlinear elements. Shown for comparison are proposed A+ and LIGO Voyager noise spectrum. \Rana{remove A+ and Voyager; instead just use DRSE with same mirror mass and arm power as WLC, lossless!}
}
\label{fig:NoiseBudget}
\end{figure}

%Let us now generalize the $\mathcal{PT}$ symmetry consideration to 
Now, for a multi-mode device with modes $\{\hat a_j\}$ coupled to signal $h$ with coefficients $\{\alpha_j\}$ and with $\sum \beta_j \hat a_j $ originally coupled to the external continuum, let us bring in systems $\{\hat c_j\}$, with $\{\hat c_j^\dagger\}$ having the same equations of motion as $\{\hat a_j\}$, and the mode $\hat b$, which couple to $\{\hat a_j\}$ and $\{\hat c_j^\dagger\}$:
\begin{align}
\dot {\hat a}_j  &=  \mathcal{M}_{jk} \hat a_k + i \alpha_j h  - 
\kappa \beta_j \hat b,\;\;
\dot {\hat c}_j^\dagger  =   \mathcal{M}_{jk} \hat c_k^\dagger  +\chi \beta_j \hat b, \\
\dot {\hat b} &= -\gamma_R\hat  b + \kappa \beta_j \hat a_j + \chi \beta_j \hat c^\dagger_j +\sqrt{2\gamma_R} \hat u\,,\;\; \hat v=\hat u-\sqrt{2\gamma_R} \hat b.
\end{align}
These equations can be solved in a matrix form and lead to:
%\begin{equation}
%\dot b + \gamma b -  \Delta  b = g\beta_k %\Lambda_{kl}\alpha_l h +\sqrt{2\gamma} u_{\xi=0-}
%\end{equation}
\begin{equation}
    \hat v=\left(\frac{\Omega-i\gamma_R-i\Delta}{\Omega+i\gamma_R-i\Delta}\right)\hat u-i\sqrt{2\gamma_R} \kappa\left(\frac{\sum_{k,l}\beta_{k}G_{kl}\alpha_l}{-i\Omega+\gamma_R-\Delta}\right)h,
\end{equation}
where
$
\Delta =
(\chi^2 -\kappa^2)\sum_{k,l}\beta_k G_{kl} \beta_l$,
$\mathbf{G} = (-i\Omega \mathbf{I}-\mathcal{M})^{-1}
$. Similar to the single-mode case,  $\mathcal{PT}$-symmetry at $\chi=\kappa$ allows $\Delta=0$, hence the cancellation of inertia existed for $\{\hat a_j\}$, and leads to a broadband amplification of signal: noise remains white while signal transfer function is proportional to $\kappa$.

%\begin{figure}
%\includegraphics[width=0.8\columnwidth]{config.pdf}
%\caption{Quantum amplifier configuration for laser interferometer GW detectors.
 % Shown in the diagram is the stable configuraion, while for unstable configuration signal
  %should be extracted from the folding mirror. \Xiang{(Xiang: should we change the figure with hatted operators? Also, the "folding mirror" should be indicated in the figure)}}
%\label{fig:config}
%\end{figure}

\noindent {\it Application to laser interferometer GW detectors --} The sWLC  can be implemented in a GW detector by making a simple modification of the uWLC: changing the location of signal extraction, as shown in Fig.~\ref{fig:uWLC}. To incorporate motions of mirrors under radiation pressure, we need to modify the interaction Hamiltonian to include $\hat x$ and $\hat p$ of the differential mode of mirror motion~\cite{kimble2001conversion}:
\begin{equation}
\hat V_{\rm GW} = -\alpha_{\rm GW} (\hat x - L h) \hat a_1  + \hat p^2/(2\mu)
\end{equation}
Here $\alpha_{\rm GW}= \sqrt{2P_c\hbar\omega_0/(Lc)}= \sqrt{\mathcal{E} \hbar\omega_0}$, with $P_c$ the circulating power in the arm ($\mathcal{E}$ is the energy stored in each arm), $\omega_0$ is the carrier angular frequency, $L$ is the arm length, and $\mu = M/4$ is the reduced mass of the differential mode ($M$ is the mass of each  mirror).  
In Fig.~\ref{fig:NoiseBudget}, we plot quantum noise spectrum (including radiation-pressure noise) for sWLC and uWLC configurations with $\gamma_R/(2\pi)=500$\,Hz,  $(\kappa,\chi)/\gamma_R=(10,9.86)$. %as well as uWLC with $\kappa=\chi=2\pi\times 10\,$kHz and $\gamma_R=2\pi\times 10\,$Hz. 
We use the LIGO Voyager~\cite{VoyInst:2020} parameters $M = 200\,$kg, $L = 4\,$km, circulating power of 3\,MW and a 2\,$\mu$m laser wavelength. We did not include the effect of optical losses, but did consider thermal noise of the $\hat c$ oscillator mode [Eq.~\eqref{eq:loss}], with  $Q_m \sim 8\times 10^9$ and $T = 4\,$K\,\cite{ma2014narrowing,miao2015enhancing}, which is the most serious source of decoherence in our scheme.

%\Xiang{%{\sl Xiang will add some more realistic parameters.}
%A more concrete numerical simulation of noise spectrum beyond single-mode approximation and with physical parameters (i.e. mirror transmittance) can be found in Supplemental Materials. More details of the numerical simulation can be found in Ref.\,(cite something in bib).}

Our proposal may also be realizable by inserting a nonlinear crystal into the signal recycling cavity, and pump with the sum frequency of the carrier frequency and another mode of the SR cavity;
this can also be viewed as a non-degenerate quantum expander~\cite{korobko2019quantum,adya2020quantum}.
%\Xiang{(Xiang:link with phase-insensitive amplification?)}
Furthermore, if we demand a full  $\mathcal{PT}$ symmetry to include the movable mirrors, we need to couple to an additional mode $(\hat X,\hat P)$ that has a {\it negative effective mass} with Hamiltonian
$    \hat V_{\rm BAE} =-\alpha_{\rm GW}\hat X \hat c_1 -\hat P^2/(2\mu)
$ analogous to the proposal of Ref.~\cite{tsang2010coherent}. We shall leave the details of this for future work, but do show their potentially achievable sensitivity in Fig.~\ref{fig:NoiseBudget}, in magenta and purple curves;  aside from quantum noise, we have only included $\hat c$-oscillator thermal noise. This significant potential for improvement motivates further studies into this direction~\cite{bentley2020systematic,shimazu2019quantum}.

\begin{figure}
\includegraphics[width=\columnwidth]{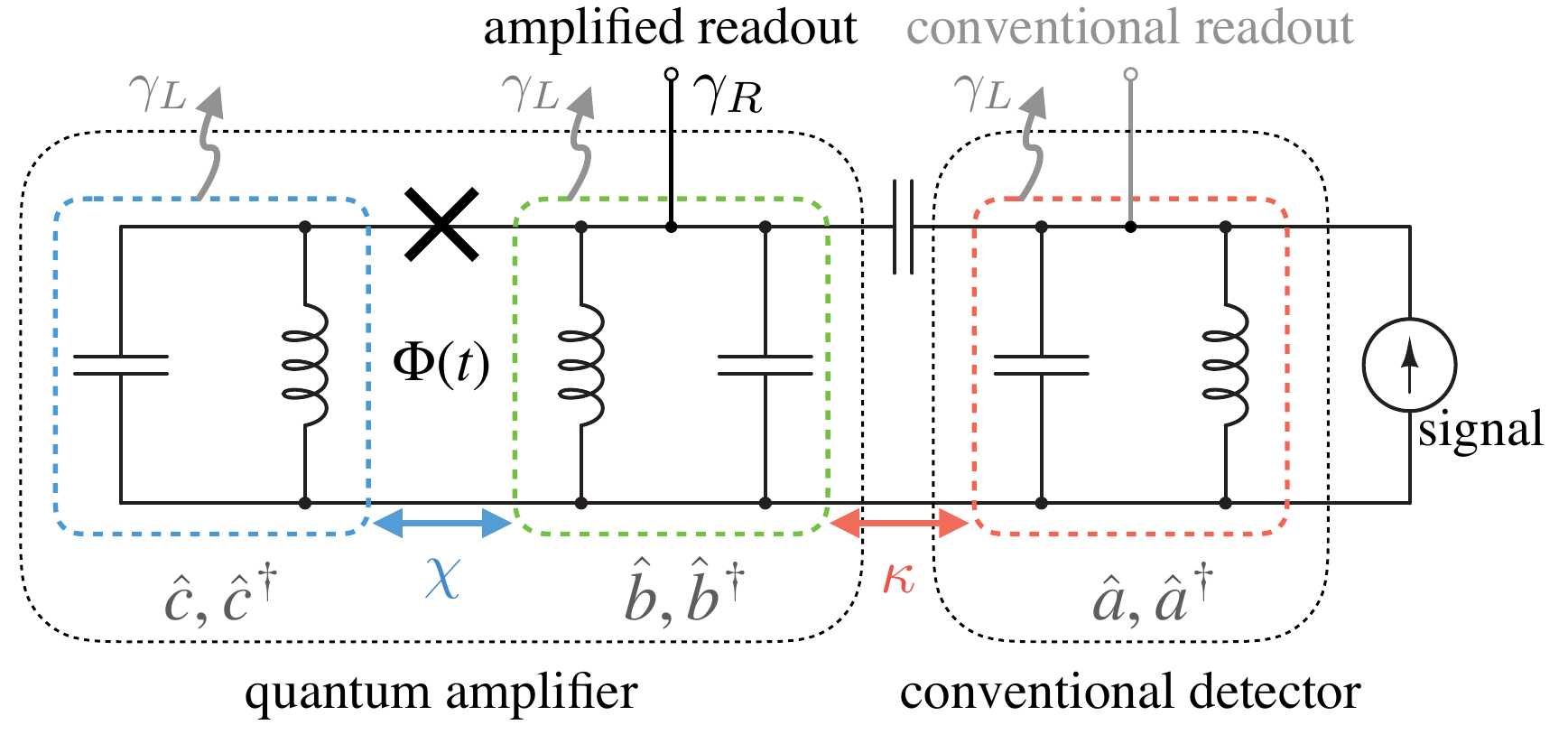}
\caption{Circuit diagram of an axion detector (dashed box on the right) attached to a quantum amplifier (dashed box on the left). The parametric amplifier is achieved by Josephson 3-wave mixer indicated by the $\times$ symbol in the figure, with $\Phi(t)$ indicating time-depending external flux that act as parametric pumping (see Fig.\,9 of Ref.\,\cite{roy2016introduction}). Coupling ``amplified readout'' to a continuum results in a sWLC, while keeping ``conventional readout'' coupling to a continuum will result in uWLC.}
%\Rana{add a Bode plot of circuit; make analogy to opamps vis-a-vis stability, include conventional and also total with readout at b and a}
%\yanbei{Rana will redraw the figure, indicating the coupling $\kappa$ (coupling between $\hat a$ and $\hat  b$), $\chi$ (coupling between $\hat  b$ and $\hat  c$), $\gamma_R$ (coupling strength to readout) and $\gamma_L$ (loss). Bode plot should contain 5 traces.  ``Conventiona'', 2 for ``stable'' and 2 for ``unstable''.  The ``stable'' transfer function is given by Eq.~\eqref{eq:v2amp}, while ``unstable'' transfer function is given by Eq.~\eqref{uWLCsol}.  }
\label{fig:axion}
\end{figure}

\noindent {\it Application to microwave axion detectors.--}
For an axion detector consisting of a stationary magnetic field $\mathbf{B}_0$ and a single mode of a microwave resonator, the interaction Hamiltonian can be written as
$
\hat V_{\rm axion} =\alpha_{\rm axion} (\Psi_1 \hat a_1+\Psi_2 \hat a_2)
$ where $\hat a_{1,2}$ are the mode quadratures, and $\Psi_{1,2}$ are two quadratures of the oscillations of the axion field %\Xiang{[? or device central frequency?]},
$\mathcal{A}(t) =\Psi(t)e^{-i\omega_0t} +\Psi^*(t) e^{i\omega_0t}$
with $\Psi_1=(\Psi+\Psi^*)/\sqrt{2}$, $\Psi_2=(\Psi-\Psi^*)/(\sqrt{2}i)$.
Here $\omega_0$ is the central frequency at which the device operates; it may need to be scanned over different values to search for the yet unknown Compton frequency of axion.

%The improvement of quantum noise will be the same as discussed earlier in this paper, with cases $g = 10\gamma$ illustrated in Fig.~\ref{fig:curves}.
%We also note that the amplification strategy we propose here can be considered a quantum version of Refs.~\cite{goryachev2019axion,goryachev2019probing}
%\textcolor{red}{Maxim and Mike? Rana?}.
%\textcolor{red}{What if we propose this for the ADMX experiment?  Does this work in their parameter regime?  Will their amplifier be quantum enough? How is this related to their existing amplifier?  }

\begin{figure}
\includegraphics[width=\columnwidth]{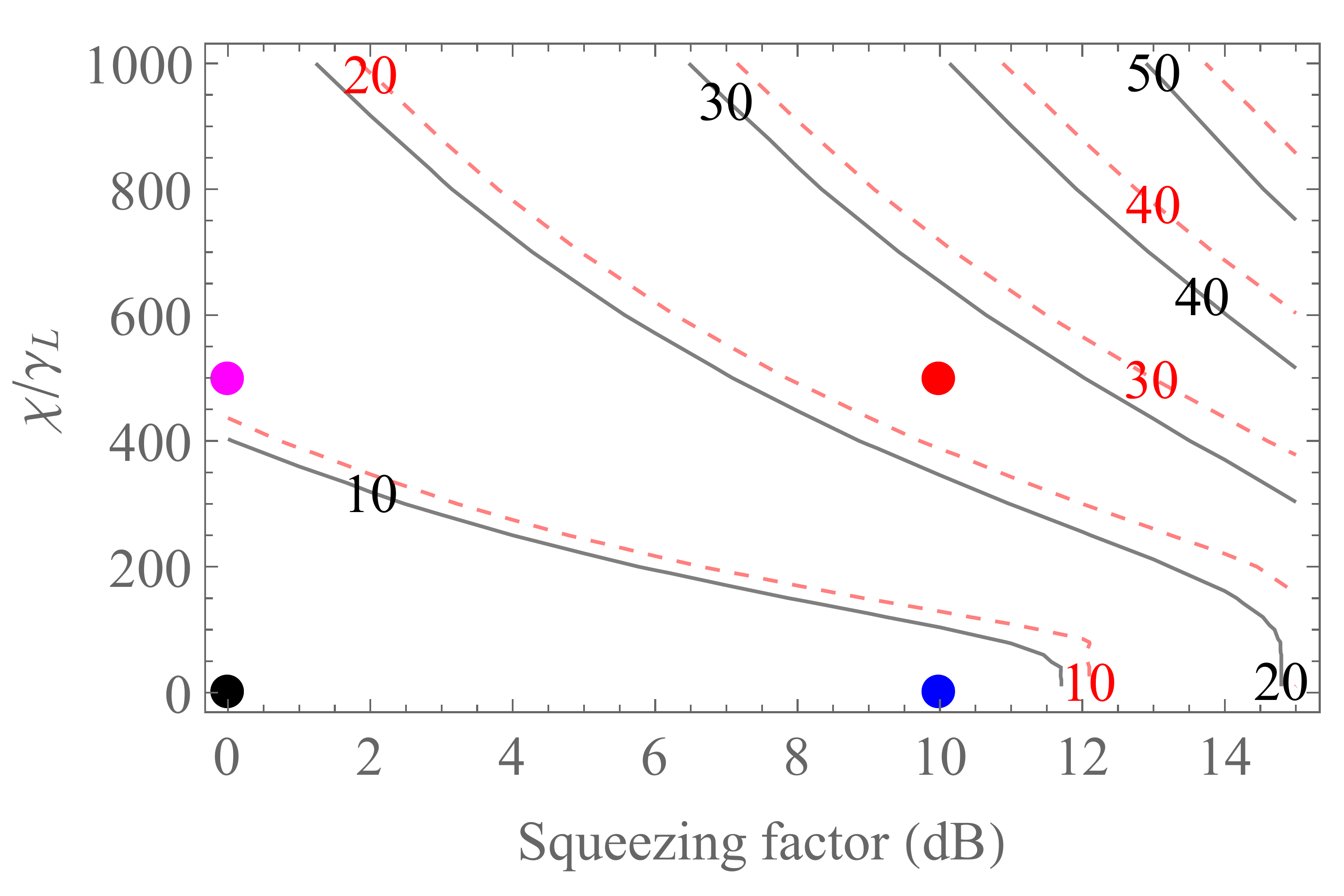}
\caption{The enhancement of the scan rate $\mathcal{R}_a$~\eqref{eq:axionFoM} achievable by sWLC axion detectors over the single-cavity one, as a function of the amplifier gain $\chi$ and squeeze factor $e^{-2r}$, optimized over the rate of $\hat a$-$\hat b$-mode coupling $\kappa$ and readout coupling $\gamma_R$. The uWLC cases optimized with the same procedure are indicated by dashed red lines for comparison, showing comparable but slightly less improvements over the single-cavity configuration. The single-cavity and sWLC configurations shown in Fig.\,\ref{fig:AXSNR} are indicated by the correspondingly colored dots.
%\yanbei{Label should be $G/\gamma_L$.  Please also mark the several configurations shown in the next plot.}
}
\label{fig:AXFoM_contour} 
\end{figure}

%\begin{figure}[ht]
%\includegraphics[width=\columnwidth]{AXFoM_dB.pdf}
%\caption{\label{fig:AXFoM_dB}Optimized FoM of axion detection behaved with different levels of squeezed vacuum input, with the same setting specified in Fig.\,\ref{fig:AXFoM_contour}.}
%\end{figure}
\begin{figure}[t]
\includegraphics[width=\columnwidth]{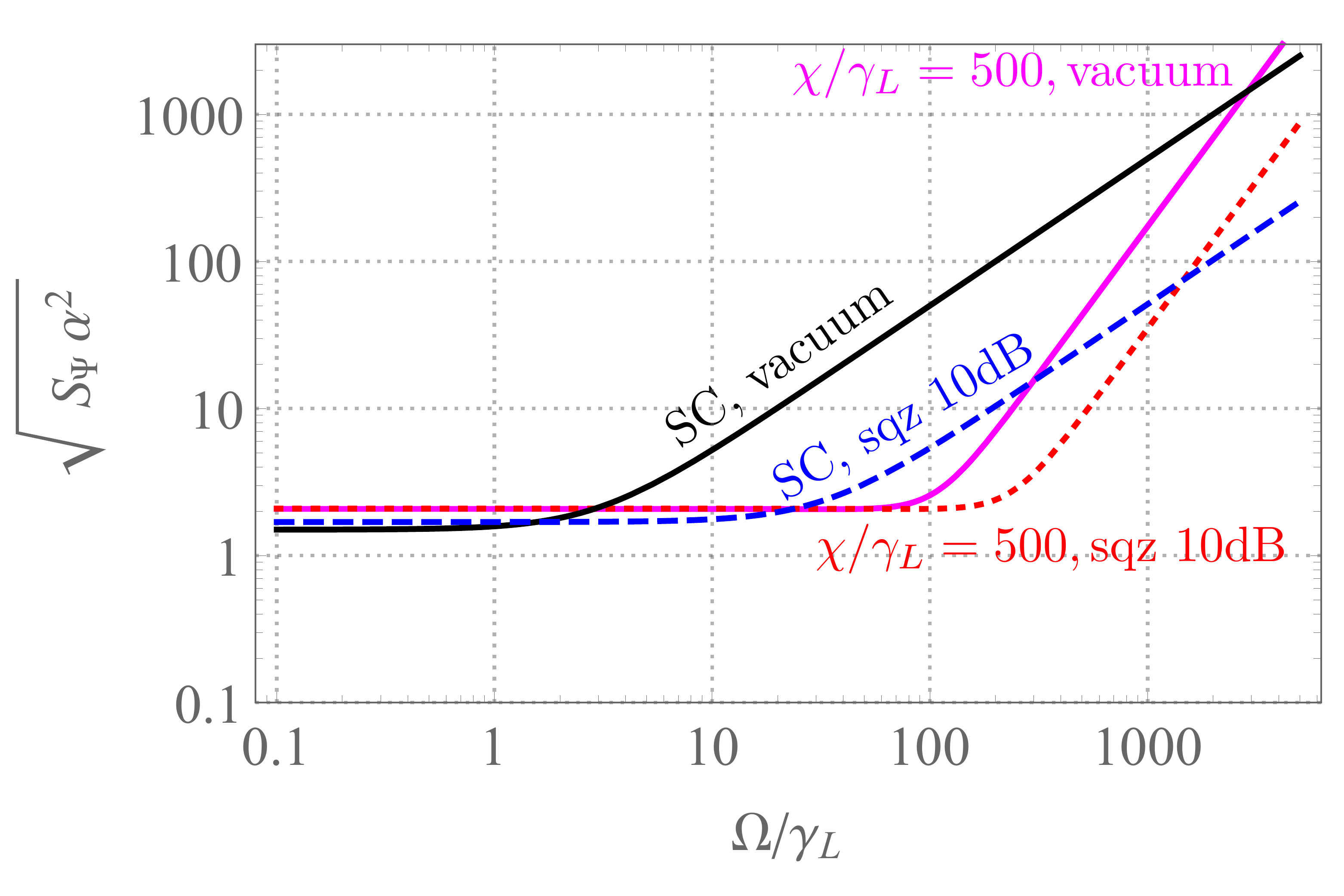}
\caption{Examples of single-cavity (SC) and sWLC noise spectrum with $\mathcal{R}_a$-optimized $\kappa,\ \gamma_R$ for different $\chi$ and input squeezing levels (sqz) $e^{-2r}$ shown by color dots in Fig.\,\ref{fig:AXFoM_contour}. See the main text for more details. %\yanbei{Please change $\Omega$ to $\Omega/\gamma_L$ and SNR to $S_\Psi$. } 
%\yanbei{Change to configurations in the middle of the contour plot. Please change $G=100$ to $\chi/\gamma_L =100$. }
}
\label{fig:AXSNR}
\end{figure}

%\noindent {\it Combination with squeezing injection.--}
A proposed  WLC circuit that realize the Hamiltonian~\eqref{eq:vamp} is shown in Fig.~\ref{fig:axion}. The parametric amplification can be realized by a 3-wave mixing device~\cite{roy2016introduction,abdo2013nondegenerate}; sWLC corresponds to having $\hat b$ mode coupled to the external continuum (``amplified readout'' in the figure), while uWLC corresponds to having $\hat a$ mode coupled to the external continuum (``conventional readout'' in the figure).   We will consider one of the axion-field quadrature $\Psi_1$, which is read out from $v_2$, and account for squeezing injection by $S_{\hat u_2 \hat u_2}=e^{-2r}$, where $S_{\hat u_2 \hat u_2}$ is the spectral density of the phase quadrature of the input field, as in Eq.~\eqref{eq:spectrumwlc}. We also assume the same loss rate $\gamma_L$ for each of the $\hat a$, $\hat b$ and $\hat c$ modes, assuming vacuum noise from each bath, and propagate these noise contributions to the output by applying Eq.~\eqref{eq:loss} to Eq.~\eqref{eq:Heqqamp} ($h$ should now be replaced by $\Psi_1$), leading to $\Psi_1$-referred noise spectrum $S_\Psi$.

%~\footnote{ In comparison with Ref.~\cite{malnou2019squeezed}, their $\kappa_l$ corresponds to our $\gamma_L$. Our $\chi$ describes coupling strength of the amplifier, and has dimension of frequency, while their $G_s$ denotes the dimensionless squeeze factor. \Xiang{(now we don't need to clarify about the same letter $G$ as we have already replaced it)}}

As a figure of merit for axion detection, we shall use the effective scan rate \cite{malnou2019squeezed,LehnertAnalysis:2020}:
\begin{equation}
  \mathcal{R}_a \equiv   \int_0^{+\infty} d\Omega/(2\pi) S_\Psi^{-2}(\Omega),
 \label{eq:axionFoM}
\end{equation}
which is inversely proportional to the search time it takes to detect the axion field, assuming the field's coherence time to be very short compared with observation time, {\it and} assuming no additional time cost when switching between configurations centered at different frequencies. In comparison with the EQL~\eqref{eq:GBWproduct}, $\mathcal{R}_a$~\eqref{eq:axionFoM} favors peak sensitivity more than bandwidth. {Considering the losses, numerical optimization is carried out on the scan rate $\mathcal{R}_a$ enhancement:} for values of $\chi/\gamma_L$ and $e^{-2r}$, search over $\kappa/\gamma_L$ and $\gamma_R/\gamma_L$ to achieve an optimal $\mathcal{R}_a$, and compare it with that of a single-cavity detector with the same $\gamma_L$ (for which the optimal $\gamma_R/\gamma_L$ is 2~\cite{malnou2019squeezed}). 
As shown in Fig.\,\ref{fig:AXFoM_contour}, the {amplifier gain $\chi$} improves $\mathcal{R}_a$ substantially, and in a way compatible with squeezing. 
In Fig.~\ref{fig:AXSNR}, we plot the noise spectra of several configurations. These configurations improve the scan rate $\mathcal{R}_a$~\eqref{eq:axionFoM} by substantially broaden the detector bandwidth, thereby requiring {\it far} less switches between different central frequencies.

%for different parametric coupling strength $G$ and squeezed vacuum input. For different $G$ and squeezing, the FoM is numerically optimized over readout rate $\gamma$ and coupling strength $g$, assuming the same intrinsic losses from all three modes $\hat a,\ \hat b,\ \hat c$ with rate $\kappa$ \Xiang{(to be replaced with more realistic values for different channels)}. 

%both larger $G$ and squeezing factor can help to improve the FoM. It is shown in more details in Fig.\,\ref{fig:AXFoM_dB} that while all cases converge with the single cavity case in the extreme squeezing limit, the FoM is greatly increased by $G$ when the squeezing level of input is limited. Some 
%representative SNRs with optimized FoMs are plotted in Fig.\,\ref{fig:AXSNR}. In the realistic case with intrinsic losses, the effects of applying sWLC and squeezed input are both to expand bandwidth while remaining the same peak sensitivity, but with different high-frequency scaling. Considering practical limitations, the combination of them should be applied.

\noindent {\it Conclusions.---}
We have proposed a quantum amplifier that can be attached to an existing sensor as a coherent quantum feedback device, and improve its sensitivity-bandwidth product. The improvement is achieved via signal amplification, and therefore, is robust against readout losses.
Our sensitivity gain can be connected to $\mathcal{PT}$ symmetry: a mode $\hat b$ is coupled to our original sensor modes $\{\hat a_j\}$, as well as auxiliary modes $\{\hat c_j\}$, which is time reversal to the original sensor modes.
Such a strategy can be used to substantially improve the sensitivity of laser interferometer gravitational-wave detectors, and microwave axion detectors.
We analyzed different figures of merit (Eqs.~\eqref{gaindef} \& \eqref{eq:axionFoM}) for the two detector schemes, while a more complete PT-symmetric structure for backaction evasion requires further study.
In previous studies of $\mathcal{PT}$ symmetry, the balance of gains and losses are often emphasized, usually in the context of non-Hermitian Hamiltonians that are constructed by eliminating external degrees of freedom \cite{ruter2010observation,peng2014parity}.
In our case, the $(\{\hat a_j\},\hat b)$ subsystem has loss, the $(\{\hat c_j\},\hat b)$ subsystem has (the balancing) gain, and the entire system has a Hermitian Hamiltonian. In this way, we are able to arrive at features similar to previous work without having to resort to external degrees of freedom. %\Xiang{(I feel like the conclusion need to include the advantage of sWLC more explicitly)}

%More broadly speaking, in the design of  classical sensors, an external feedback controller can often be used to undo the frequency-dependent phase-shift inside a {\it plant} to enhance its response to signal and hence sensitivity; we have shown that the same principle can be applied to the quantum setting, except the filtering should be done quantum mechanically using an auxiliary system that is the time-reversed version of the plant.

\noindent {\it Acknowledgments.---}
Y.C., X.L., and Y.M.'s research was funded by the US National Science Foundation, and the Simons Foundation.
Part of the this research was completed when Y.C. was visiting the OzGrav centers of Excellence at the University of Western Australia and the Australian National University. 
R.X.A. acknowledges support by the De Logi Science and Technology Trust, De Logi Science and Tech grant, as well as support under NSF-Voyager, NSF award PHY-1912677. 
M.E.T and M.G. are funded by Australian Research Council Grant Numbers,  CE170100009, CE200100008 and DP190100071.
We would like to thank Gray Rybka and Aaron Chou for illuminating discussions on axion detection, and the LIGO Quantum Noise Working Group.
%R.X.A. acknowledges support by the De Logi Science and Tech grant, as well as support under NSF-Voyager. 

\noindent
{\it Note added.--}During the completion of this work, we became aware of a related but different study\,\cite{yan2020coherent}.

% please import the BibTex entries directly from the journal
% rather than typing them in, to avoid the errors in the refs
\bibliographystyle{apsrev}
\bibliography{references}

\end{document}